\documentclass[11pt,twoside]{article}

\usepackage{asp2004}
\usepackage{epsf}
\usepackage{psfig}
\usepackage{lscape}
\usepackage{graphicx}

\begin{document}
\title{MHD Models for Winds from Neutron Stars, and their Interaction with the Environment}   
\author{Niccol\`o Bucciantini}   
\affil{Astronomy Dep., University of California, Berkeley}    

\begin{abstract} 
The recent development of numerical schemes for Relativistic MHD (RMHD) allows us to model the acceleration and outflow properties of winds from compact sources. Theoretical models suggest that acceleration and collimation of the flow are extremely inefficient when the speed is close to $c$, in contrast with many observations. Numerical results for an axisymmetric rotator, both in the case of monopolar and dipolar magnetic field will be presented, suggesting that in ideal RMHD acceleration is indeed inefficient. I will also mention numerical challenges and stability problems of present simulations. Finally I will discuss the interaction of a pulsar wind with the surrounding SNR. I will present emission maps based on numerical simulations of the flow inside the nebula and how they can be used in comparison with observations to derive informations about the properties of pulsar winds far away from the source.
\end{abstract}


\section{Introduction}   

Magnetically dominated outflows from stars and accretion discs, are central to the  evolution and properties of these objects. Because they can efficiently extract rotational energy, and convert it into wind kinetic energy, magnetic outflows are ubiquitous in powering a variety of astrophysical systems. Rotationally powered pulsars are among the best example of relativistic outflows produced by a fast compact rotator. The high magnetic field at the surface ($B\sim 10^{10-13}$ G) and short rotation period ($P\sim 0.001-1$ s), induce an electric field at the surface itself which is strong enough to lift electrons and ions and accelerate them to relativistic speeds. These accelerated particles, as they move along curved magnetic field lines, produce the pulsed emission that is observed. In the magnetosphere of the neutron star, these particles produce pairs. If the charge density exceeds the Goldreich-Julian value then conditions can be modeled in term of ideal RMHD.

The typical Lorentz factor of the plasma in the pulsar magnetosphere, within the light cylinder, is $\gamma\sim 100$. However from observations of the interaction of the pulsar wind with the environment, and especially the surrounding SNR, we know that far away from the source the wind is ultrarelativistic ($\gamma\sim 10^{4-6}$) and weakly magnetized \citep{kc84a,kc84b,arons04}. The basic idea is that the wind is magneto-centrifugally accelerated, and that the high Lorentz factor is achieved by almost completely converting magnetic energy into kinetic energy. 

In this sense, pulsars constitute the prototype of relativistic accelerators and the basic physics of magneto-centrifugally driven relativistic winds apply to a wide variety of astrophysical objects from accretion disks to GRBs and microquasars. In all of these systems a plasma is assumed at the base with small outflow velocity (compared to the asymptotic velocity), embedded in a strong magnetic field. Magnetorotation then accelerates the plasma and converts the magnetic energy into relativistic motion. The study of winds from neutron stars can give an useful insight into the general problem of relativistic outflows.

In the simple 1D radial model for relativistic winds from compact rotators by \citet{michel69}, solutions can be parametrized in term of $\sigma=\Omega^2\Phi^2/\dot{M}$, where $\Omega$ is the rotation rate, $\Phi$ the magnetic flux and $\dot{M}$ the mass flux. $\sigma$ is the maximum Lorentz factor the wind can achieve if all magnetic energy is converted into kinetic energy, and at the base of a pulsar it is $\sim 10^6$. The results by \citet{michel69} show that, due to relativistic effects, the asymptotic Lorentz factor tends to $\sigma^{1/3}$, and conversion of magnetic to kinetic energy is minimal. This is known as the $\sigma-\gamma$ problem \citep{arons04}. 

The low efficiency of the Michel model is due to the assumption of strictly radial flow. If this assumption is relaxed and the flow is allowed to diverge more than radially from the Light Cylinder up to the SNR, then acceleration is more efficient \citep{begelman94}. It is known from the theory of classical winds that the presence of a magnetic field can cause collimation of the flow along the rotation axis, and divergence along the equator, so in principle one might expect magnetic collimation to be also responsible for the acceleration of the wind. However when the flow speed reaches values close to $c$ the Coulomb term in the electromagnetic force cannot be neglected (as in non relativistic MHD). The strength of the induced electric field is comparable with the magnetic field and the Coulomb term balances the magnetic hoop stress. Solutions of the relativistic Grad-Shafranov equation for a monopolar field in the trans-fast regime \citep{beskin98} have shown that acceleration is too slow and the asymptotic results of the Michel model apply also in 2D axisymmetric geometry, meaning that the flow remain radial. 

All simulations were performed using the code developed by \citet{ldz03}. It is a high-order finite difference UCT scheme, that solves the equation of relativistic MHD in general curvilinear coordinates. The scheme is based on an approximate HLL solver, that avoids wave decomposition.

\section{Wind acceleration}  

The first axisymmetric numerical simulations of the acceleration of a relativistic wind were presented by \citet{bog99}, where the case of a zero pressure plasma in a monopolar magnetic field, with uniform injection conditions at the base of the star was investigated. Results show that when $\gamma$ exceeds 5 collimation and acceleration of the wind become marginal. 

Our recent investigation \citep{buc06} has focused on the study of the acceleration of relativistic wind in the case the pressure is small compared with the rest mass energy density but not negligible. It has been proposed that soon after the birth of a proto neutron star (NS), neutrino heating might rise the sound speed in the neutron star atmosphere to value $\sim 0.1c$ \citep{thom04}. The heating might be able to drive a wind against the gravitational attraction of the NS. Such wind might be accelerated to high $\gamma$ by the magnetic field. If acceleration is efficient, the wind can reaches $\gamma\sim 100-300$, enough to trigger a GRB. We have investigated the axisymmetric wind from rotating neutron star in Schwartzschield metric. Pressure, density and the radial component of the magnetic field are fixed at the NS surface ($r_{NS}$). A 2D spherical $r-\theta$ grid was used, with 100 uniform cells in $\theta$, and a logarithmic scale with 200 cells per decade in $r$.

Solutions in the 2D case can be parametrized by using the value $\sigma$ averaged over all stream lines. Results in the case of a monopolar magnetic field show that for $\sigma > 1$ the flow appears to be radial, the terminal $\gamma$ is $\ll \sigma$, pointing to inefficient acceleration. Moreover the energy and angular momentum fluxes rapidly approach the force-free value. Despite the fact that in the close magnetosphere $v \ll c$, asymptotic results agree with the flow structure found by \citet{bog99}. Energy flux scales as $\sin^2{(\theta)}$, and  $\gamma\propto\sin{(\theta)}$, in agreement with the exact monopole solution. Results suggest that a rapidly rotating magnetar will not be able to simultaneously produce a collimated and relativistic outflow. Relativistic velocities can be reached only for $\sigma\gg 1$, and in this case most of the energy would be concentrated in the equatorial plane instead that in a jet.

\begin{figure}
\includegraphics[scale=0.4,clip=true]{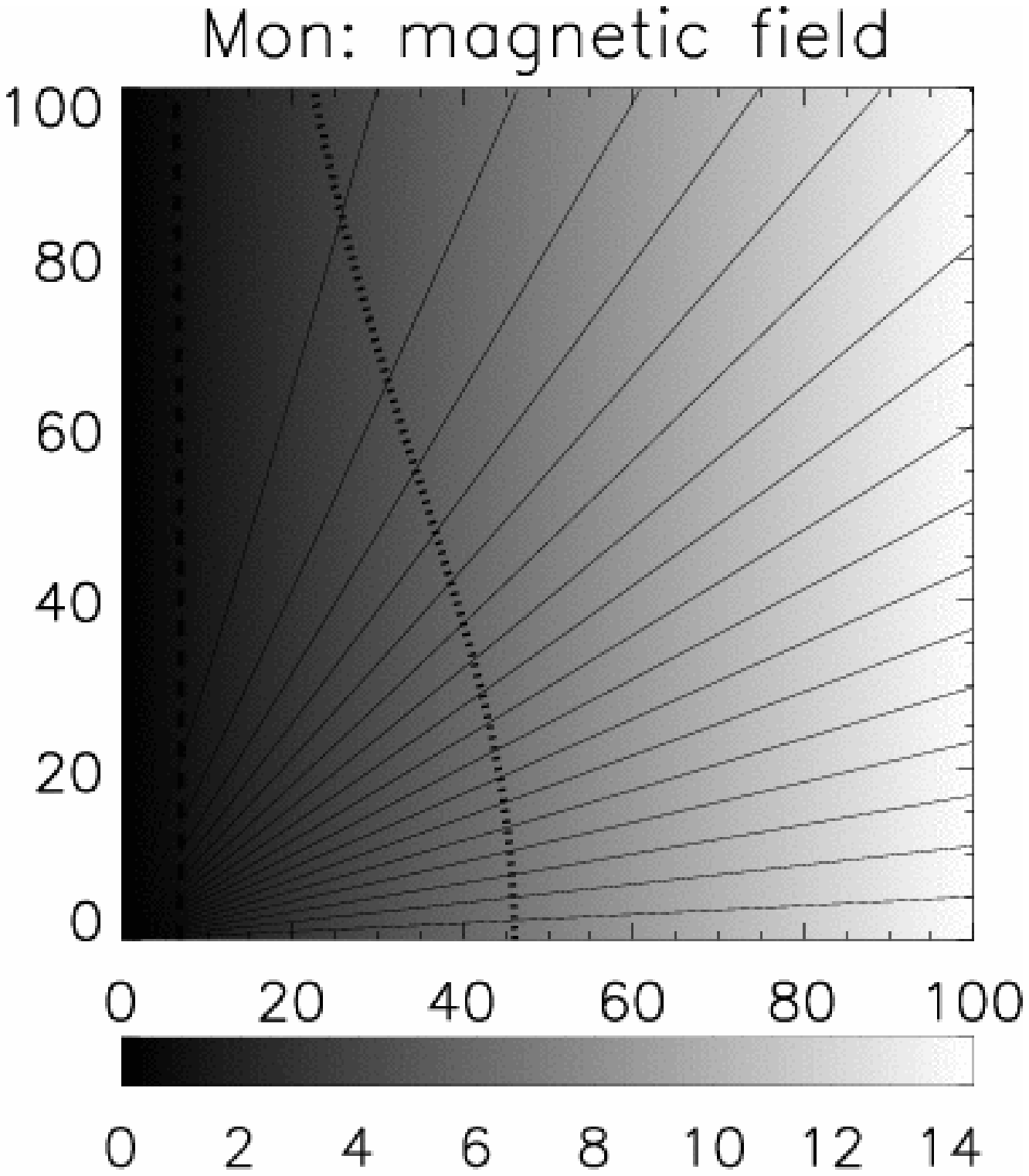}\hspace{1cm}\includegraphics[scale=0.4,clip=true]{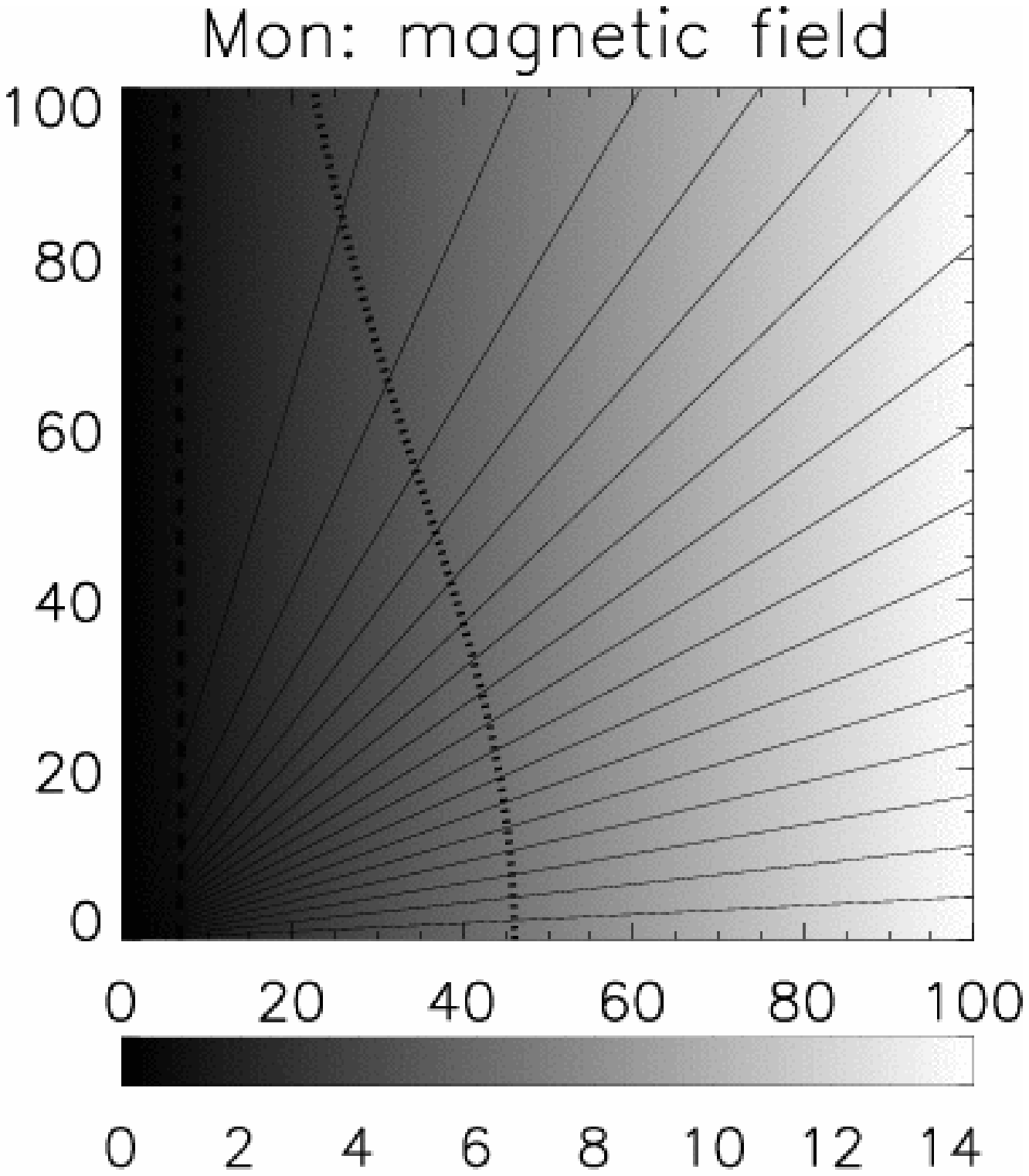}
\caption{Wind structure for a monopolar magnetic field, $\sigma=200$, axes are in units of $r_{NS}$. Left: contours represent magnetic surfaces, colors represent $|B_{\phi}/B_r|$, the dashed line is the Alfvenic surface, coincident with the light cylinder, the dotted line is the fast-magnetosonic surface. Right: Lorentz factor in the wind. From \citet{buc06}}
\end{figure}

We computed solutions also in the more realistic cases of a dipolar magnetic field at the surface of the NS. Due to stability issue in numerical RMHD it was not possible to study cases with $\sigma > 20$. Numerical resistivity on  current sheets at the edge of the closed zone, causes the plasma to rotate faster that corotation, while at the equator reconnection takes place and plasmoids are formed. To obtain steady solutions one is forced to suppress resistivity, thus enhancing the instability of the code. Moreover, in the case of a dipole the magnetic and flow surface are not aligned with the grid, and in this case inversion from conservative to primitive variables is not stable for $B^2/\rho c^2 > 100$.    

Despite the presence of a closed zone, and the fact that within the light cylinder the field structure is close to a dipole, we found that solutions can again be parmaterized in term of $\sigma$, now defined as an average done only on open streamlines. In this case we find that energy and angular momentum losses are the same of a monopolar case with the same $\sigma$. Again as soon as $\sigma> 1$ solutions approach the force-free limit \citep{cont99}, outside the Light Cylinder, and for $\sigma >10$ the asymptotic structure resembles the exact monopole, with energy flux peaking at the equator and a higher equatorial Lorentz factor. Similar results are found also by \citet{kom06}. 

Contrary to expectations \citep{cont99} we also find that the location of the Y-point is inside the Light Cylinder. This is due to the presence of a finite pressure plasma at the base of the star. In this case the extent of the closed zone is limited by the pressure equilibrium at the Y-point. This implies that even for large $\sigma$ the asymptotic wind structure is close to the force-free monopole solution, the spindown time can be shorter, and the total energy losses larger that what is inferred by applying the dipole formula. 

\begin{figure}  
\includegraphics[scale=0.4]{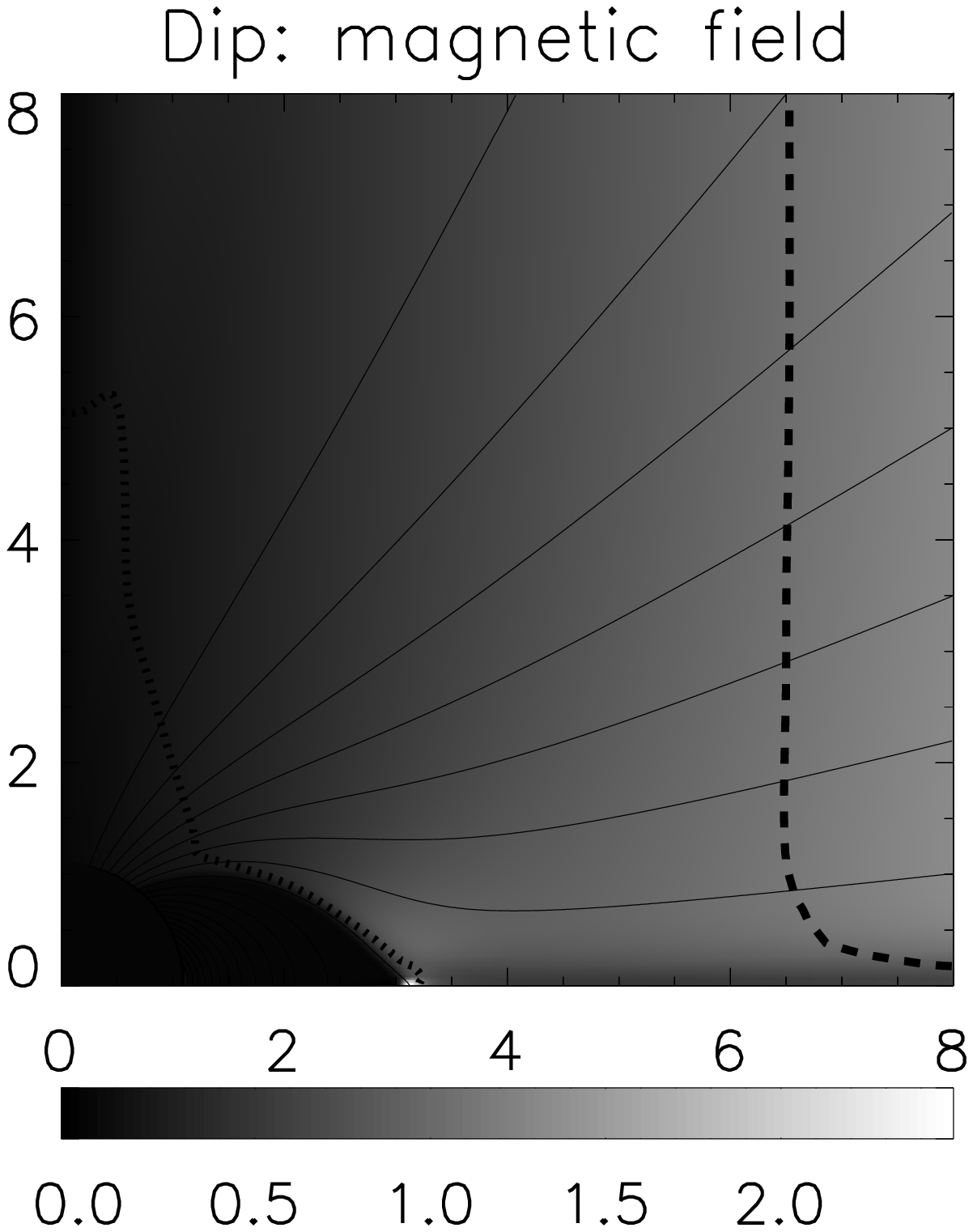}\hspace{1cm}\includegraphics[scale=0.4]{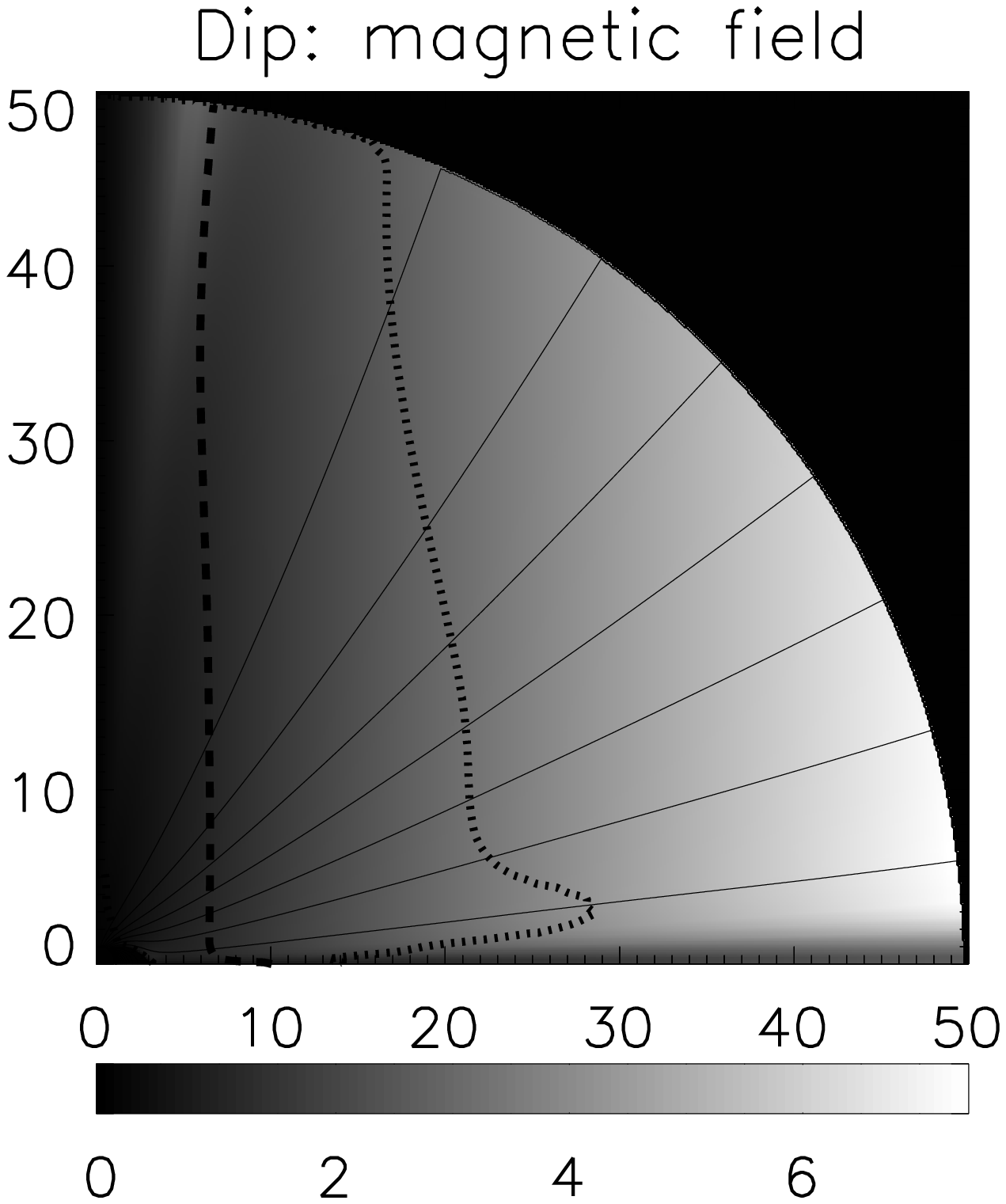}
\caption{Wind structure for a dipolar magnetic field, $\sigma=20$, axes are in units of $r_{NS}$. Left: contours represent magnetic surfaces, colors represent $|B_{\phi}/B_r|$, the dashed line is the Alfvenic surface, coincident with the light cylinder, the dotted line is the slow-magnetosonic surface. Right: same as the left panel, the dotted line is now the fast-magnetosonic surface. Note the radial structure in the asymptotic region. From \citet{buc06}.}
\end{figure}

\section{Pulsar Wind Nebulae}  

A way to constraint the properties of the wind at large distances is to study its interaction with the environment and especially with the surrounding SNR. The pulsar wind blows a cavity inside the SNR, filled with a hot relativistic pair plasma and magnetic field, known as Pulsar Wind Nebula, which shines in non thermal synchrotron emission from radio to X-rays. Recent X-ray images of the central region in PWNe have shown the presence of a well defined axisymmetric structure known as ``jet-torus''. A main emission torus is observed in the equatorial plane, with the presence of an inner ring of emission; a perpendicular jet is present. Sometimes an enhancement of emission close to the pulsar location, known as the ``knot'' is also observed. The observed doppler boosting suggests velocities in the torus $\sim 0.3-0.4c$ and in the jet $\sim 0.6-0.7c$.  Such structure is too complex to be interpreted in term of existing 1D models. Moreover the presence of a jet seems to contradict the result on the wind structure that collimation is not efficient. 

Recent numerical simulation \citep{ldz04,ldz06} has shown that such structure naturally arises as a consequence of an anisotropic energy distribution and sufficiently high magnetic field in the nebula. Assuming a split monopole for the wind energy and lorentz factor distribution, as derived from 2D simulation of relativistic winds, and a magnetization higher than $\sigma\sim 0.01$, simulated synchrotron maps are able to recover qualitatively all the main features of the jet-torus structure. The torus and inner ring as well as the knot are explained in term of region where the flow is moving at a relativistic speed toward the observer, while the formation of the jet is associated with collimation by hoop stresses inside the nebula.   In all simulations a 2D spherical $r-\theta$ grid was used, with 100 uniform cells in $\theta$, and a logarithmic scale with 200 cells per decade in $r$. A wind with energy distribution typical of a monopole is injected at the base, in an environment representative of SNR ejecta.

\begin{figure}
\includegraphics[scale=0.4]{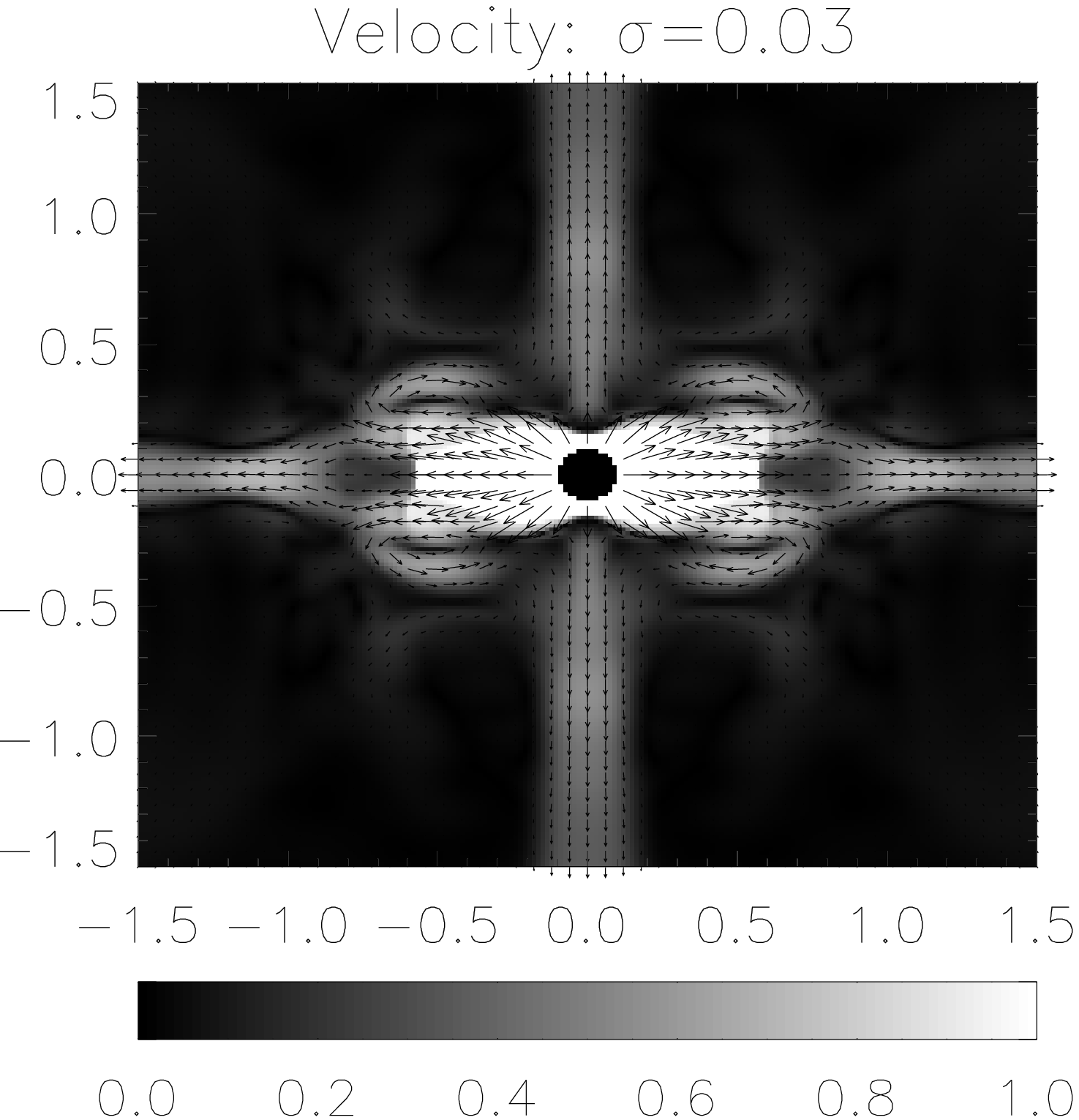}\hspace{1cm}\includegraphics[scale=0.4]{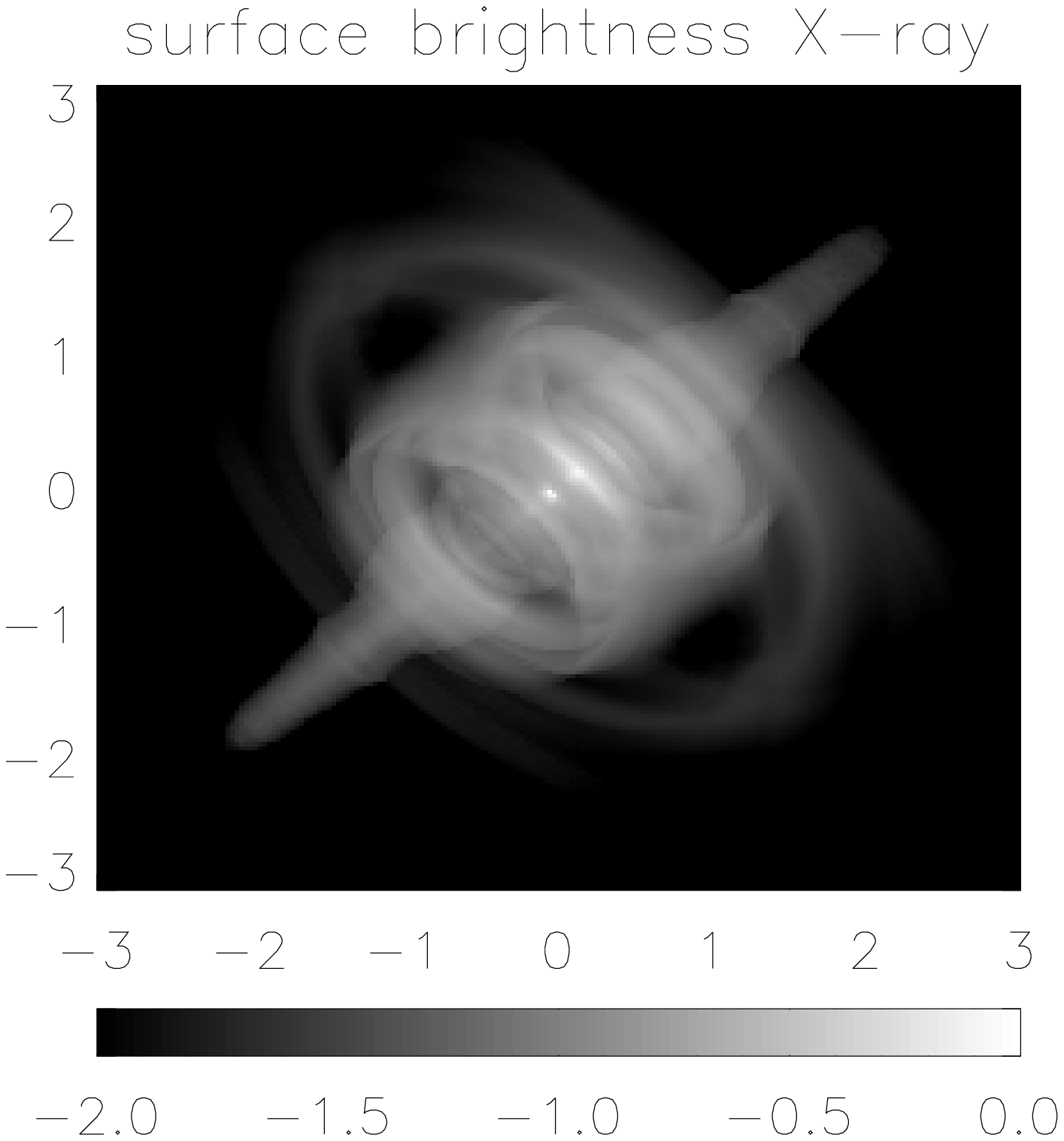}
\caption{Right: simulation of the flow structure inside a PWN. The oblate shape of the wind region is due to the wind energy distribution. A fast equatorial flow channel survives because magnetization vanishes at the equator. At higher latitudes the flow is collimated toward the axis by hoop stresses and forms a jet. From \citet{ldz04}. Left: synchrotron map based on simulations. Notice the presence of a central ``knot'' an inner arc and an outer torus. The jet is also visible. From \citet{ldz06}. Axes in both pictures are in light years.}
\end{figure}

Results suggest that the emission features observed in the nebulae might be  compared with simulations to constraint the wind properties. Preliminary results show that a better agreement is achieved if an unmagnetized region in the equatorial plain is assumed. Such unmagnetized region could be due to dissipation  of the current sheet \citep{kirk03}, which in turn could provide a mechanism for accelerating the wind and solving the $\sigma$ problem.

\section{Conclusions}

Recent numerical results have confirmed theoretical predictions about the acceleration of relativistic pulsar winds. The conversion of magnetic to kinetic energy appears to be minimal, the flow closely resembles the exact monopole solution, there is no evidence for the formation of a collimated energetic jet. The use of a more realistic dipolar field does not seem to change significantly the wind properties far from the light cylinder. On the contrary the energy and angular momentum losses seem to depend only on the integrated value of $\sigma$, and for $\sigma>1$ on the amount of open magnetic flux.

An alternative way to investigate the wind properties at least at large distances comes from the study of its interaction with the surrounding SNR. Modeling of the ``jet-torus'' structure in PWNe have shown that  such features are due to the details of the nebular flow, which in turn depends on the wind properties. Simulations show that an energy and magnetic field distribution in the wind resembling the monopole solution, can indeed explain the main details of the observations. 


\acknowledgements 
This work has been supported by NSF grant AST-0507813, and by NASA grant NAG5-12031. 


\end{document}